\documentclass[5p]{elsarticle}

\usepackage{amsmath}
\usepackage{lineno,hyperref}
\usepackage{makecell}
\usepackage{threeparttable}
\hypersetup{colorlinks = true}
\modulolinenumbers[5]

\journal{Journal of \LaTeX\ Templates}

\usepackage[some]{background}
\SetBgScale{1}
\SetBgContents{\parbox{10cm}{%
  \Huge Draft:  \today\\[18cm]\rotatebox{180}{\Huge Draft:  \today}}}
\SetBgColor{red}
\SetBgAngle{270}
\SetBgOpacity{0.2}

\biboptions{authoryear}

\begin{document}

\begin{frontmatter}


\title{Digital Twins for Marine Operations: A Brief Review on Their Implementation}

\author{Federico Zocco\fnref{fn1}}

\author{Hsueh-Cheng Wang\fnref{fn2}} 

\author{and Mien Van\corref{mycorrespondingauthor}\fnref{fn1}}
\fntext[fn1]{Centre for Intelligent Autonomous Manufacturing Systems, School of Electronics, Electrical Engineering and Computer Science, Queen's University Belfast, Northern Ireland, UK. \\ Email: federico.zocco.fz@gmail.com, m.van@qub.ac.uk}

\fntext[fn2]{Department of Electrical and Computer Engineering, National Chiao Tung University (NCTU), also with the Institute of Electrical and Control Engineering, and National Yang Ming Chiao Tung University (NYCU), and also with the Pervasive Artificial Intelligence Research (PAIR) Labs, Hsinchu, Taiwan. \\ Email: hchengwang@g2.nctu.edu.tw}

\cortext[mycorrespondingauthor]{Corresponding author}

\begin{abstract}
While the concept of a digital twin to support maritime operations is gaining attention for predictive maintenance, real-time monitoring, control, and overall process optimization, clarity on its implementation is missing in the literature. Therefore, in this review we show how different authors implemented their digital twins, discuss our findings, and finally give insights on future research directions.   
\end{abstract}


\end{frontmatter}


\section{Introduction}
System modeling is an approach used in engineering to evaluate the behavior of a system for different working conditions. Without models, it would be necessary to create the working conditions on the real system, which is usually economically and time expensive. In addition, a model can also be used to design feedback controllers that guarantee certain requirements such as safety, stability, and optimality, which would be prohibitive to do directly on the real system. Clearly, this is possible only if the model provides an accurate approximation of the relationship between the physical variables of interest, otherwise any insights provided by the model are of little relevance for the real system operation.

Ideally, the model should be a \emph{twin} of the real system, so that any change and analysis of the real system can be performed on the digital copy within short time and with minimal economic costs. This is the key idea of \emph{digital twins} (DT) (\cite{grieves2017digital,jones2020characterising,singh2021digital}). Despite the concept of digital twins dates back to 2002 (\cite{grieves2017digital}), it is currently missing in the literature a review of how different authors have implemented them for marine operations. Moreover, it is sometimes unclear the difference between standard modeling and digital twins. With this review, we give the following main contributions to address the mentioned literature gaps.    
 
\begin{enumerate}
\item{While DTs are gaining attention in the marine literature, it is missing clarity on how different authors implemented them; therefore, we provide a review on implementation aspects for maritime applications.}
\item{By focusing on implementation aspects of DTs, we observe that the trend in the maritime literature is to identify modeling as the main component of a digital twin; in contrast, our review highlights how a digital twin is far more complex than standard modeling and suggests future research direction.}
\end{enumerate}

\section{Methods}\label{sec:methods}
This section details the methodology used to perform the literature review given in Section \ref{sec:results}. First of all, we adopted the following definition of a digital twin that slightly re-words the definition from the literature to emphasize implementation aspects, which are the aim of this paper.  

\begin{figure}
\includegraphics[width=0.45\textwidth]{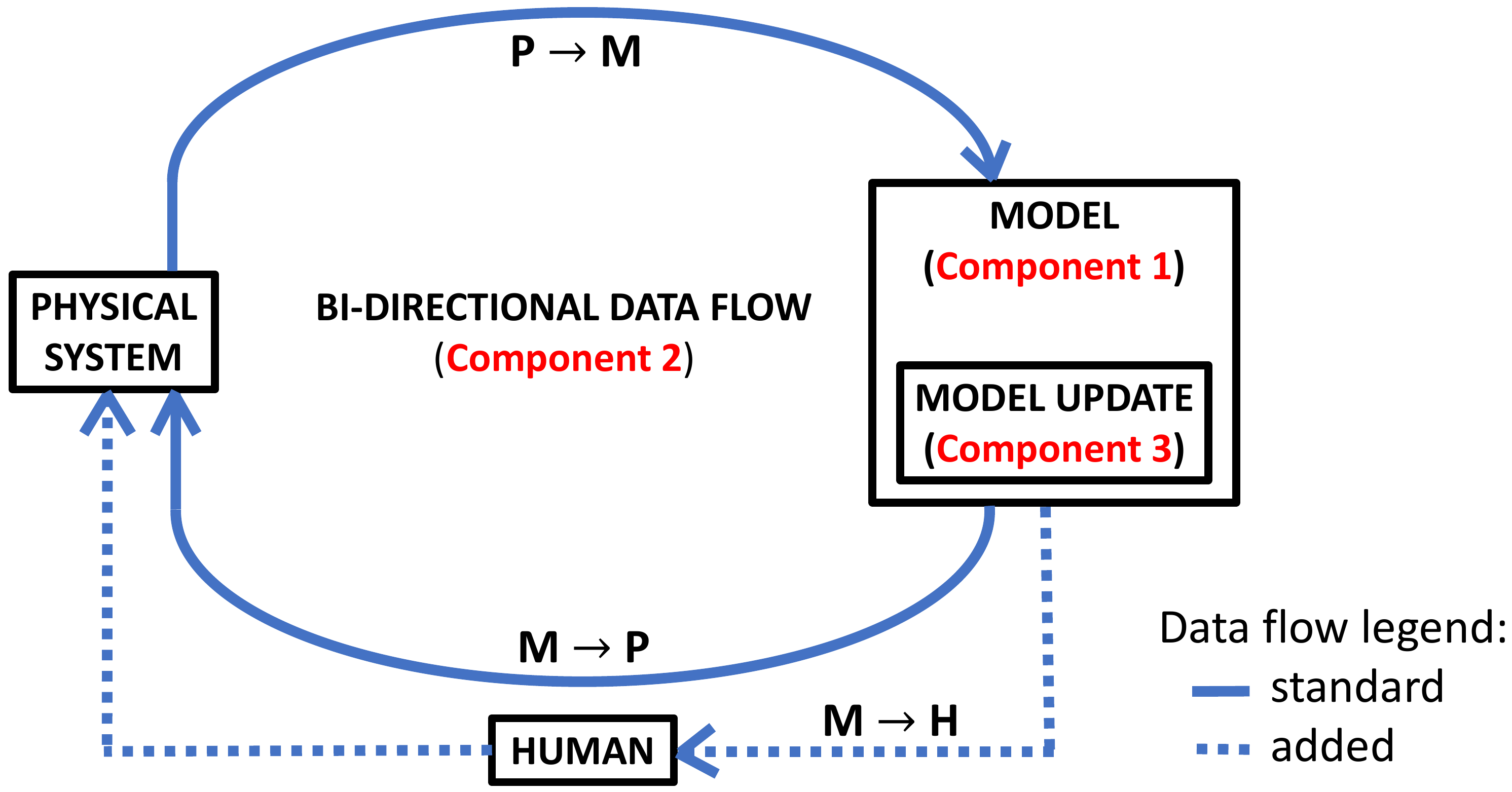}
\centering
\caption{Diagram of the main components and the data flows of a digital twin. The nomenclature used in this diagram is adopted throughout the paper.}
\label{fig:SchemeOfDT}
\end{figure}

\vspace{0.2cm}
\noindent
\textbf{Definition 1 \citep{grieves2017digital,jones2020characterising,singh2021digital}:} \emph{A digital twin (DT) is a software platform aiming at mirroring the dynamics of a physical system. A DT is made of three main components: 
\begin{enumerate}
\item{a model, i.e., Component 1;}
\item{a bi-directional flow of data between the physical system and the model, i.e., Component 2;}
\item{a model update so that the model changes over time according to the physical system, i.e., Component 3.}
\end{enumerate}}
In line with this definition, we reviewed works found by intersecting the keywords ``digital twin'' and ``maritime'' or ``digital twin'' and ``marine'' with this question in mind: which ones of the three main Components in Definition 1 are implemented by the authors? Specifically for Component 2, we also looked at which flow of data was considered in each article: throughout this paper, the flow between the physical system and the model is indicated with P $\rightarrow$ M, whereas the flow from the model to the physical system is indicated with   
M $\rightarrow$ P. While a standard digital twin does not involve a human-in-the-loop, we also consider the data flow between the model and a human because it was found in some articles. This data flow is indicated with M $\rightarrow$ H. Accordingly, a bi-directional flow between the physical system and the model is indicated with P $\leftrightarrow$ M, whereas a flow from the physical system to the model and then to the human is indicated with P $\rightarrow$ M $\rightarrow$ H. The three Components of a digital twin and the data flows are summarized in Fig. \ref{fig:SchemeOfDT}. The next subsections explain what this review considers to be a model update, i.e., Component 3, making the distinction between two different modeling approaches found in the literature: physics-based modeling and statistics-based modeling. Note that learning-based approaches such as machine learning are classified as statistics-based methods.

\subsection{Model Update in Physics-Based Modeling}   
Consider the change of a model parameter, e.g. the vehicle mass, implemented to follow the same change occurred in the physical system: is this a model update? This review considers as a model update only a modification of the relationship between the symbolic terms involved in the model, i.e., a modification of the functions. A change of the values of the mathematical symbols, i.e., parameters and variables, is not considered a model update, but rather a P $\rightarrow$ M data flow. Therefore, the answer to the question above is negative. 

\begin{figure}
\includegraphics[width=0.45\textwidth]{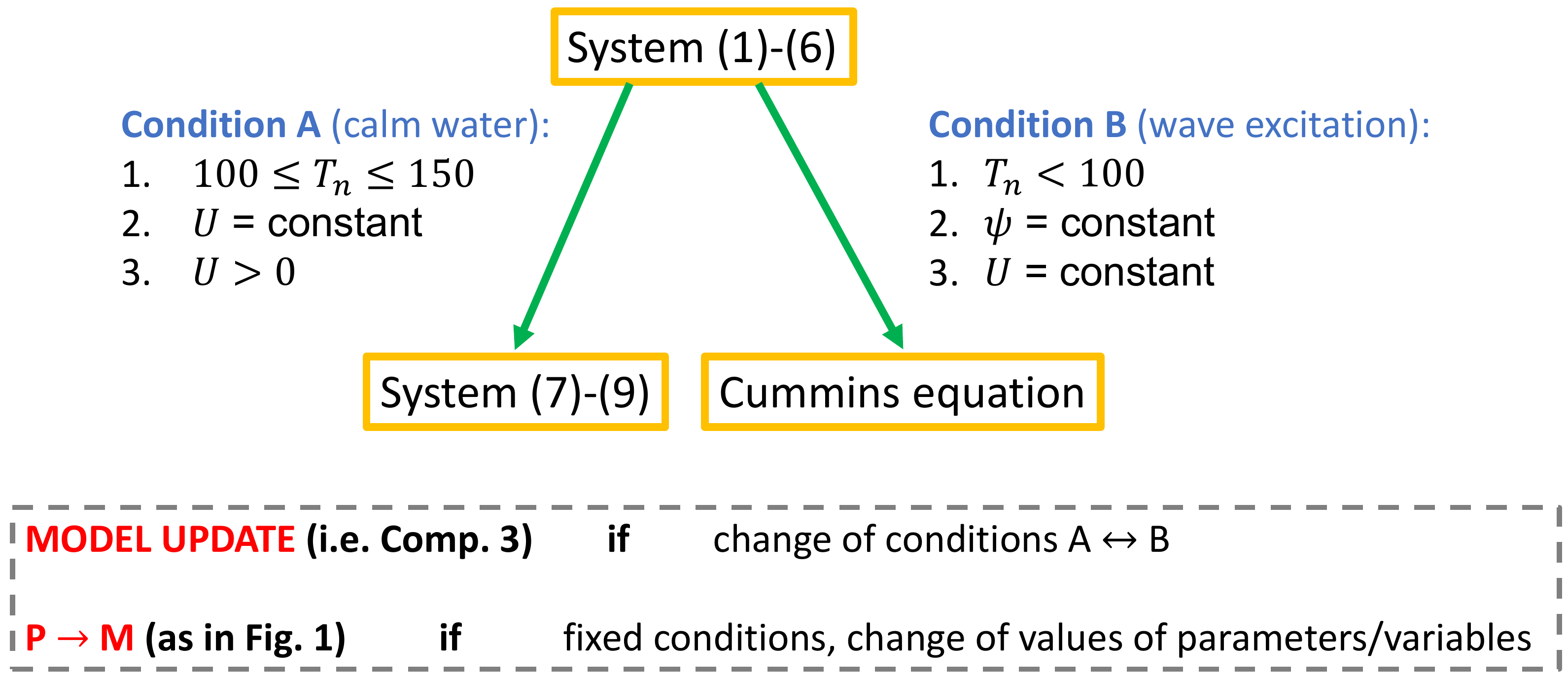}
\centering
\caption{Example of what this review considers to be a model update and a P $\rightarrow$ M data flow. This is valid if the modeling is based on physics laws rather than on statistics.}
\label{fig:ModelUpdateExample}
\end{figure}

As a further example, consider the digital twin of a ship. Under the assumption of a rigid body, its dynamics is described by the system of six differential equations (\ref{eq:6DOF-eq1})-(\ref{eq:6DOF-eq6}) \citep{fossen2011handbook},
\begin{figure*}
\begin{align}
m[\dot{u} - vr + wq - x_{g}(q^2 + r^2) + y_{g}(pq - \dot{r}) + z_{g}(pr + \dot{q})] = X \label{eq:6DOF-eq1}\\
m[\dot{v} - wp + ur - y_{g}(r^2 + p^2) + z_{g}(qr - \dot{p}) + x_{g}(qp + \dot{r})] = Y  \label{eq:6DOF-eq2}\\
m[\dot{w} - uq + vp - z_{g}(p^2 + q^2) + x_{g}(rp - \dot{q}) + y_{g}(rq + \dot{p})] = Z \label{eq:6DOF-eq3}\\
I_{x}\dot{p} + (I_{z} - I_{y})qr - (\dot{r} + pq)I_{xz} + (r^2 - q^2)I_{yz} + (pr - \dot{q})I_{xy} + m[y_{g}(\dot{w} - uq + vp) - z_{g}(\dot{v} - wp + ur)] = K \label{eq:6DOF-eq4}\\
I_{y}\dot{q} + (I_{x} - I_{z})rp - (\dot{p} + qr)I_{xy} + (p^2 - r^2)I_{xz} + (qp - \dot{r})I_{yz} + m [z_{g}(\dot{u} - vr + wq) - x_{g}(\dot{w} - uq + vp)] = M \label{eq:6DOF-eq5}\\
I_{z}\dot{r} + (I_{y} - I_{x})pq - (\dot{q} + rp)I_{yz} + (q^2 + p^2)I_{xy} + (rq - \dot{p})I_{xz} + m[x_{g}(\dot{v} - wp + ur) - y_{g}(\dot{u} - vr + wq)] = N \label{eq:6DOF-eq6}
\end{align}
\end{figure*}
where X, Y, Z, K, M, N are the external forces and moments, $m$ is the ship mass, $x_{g}$, $y_{g}$, $z_{g}$ are the coordinates of the center of gravity, $u$, $v$, $w$ are the linear velocities, $p$, $q$, $r$ are the angular velocities, $I_{x}$, $I_{y}$, $I_{z}$ are the moments of inertia and $I_{xy}$, $I_{yz}$, $I_{xz}$ are the products of inertia. The external forces and moments in (\ref{eq:6DOF-eq1})-(\ref{eq:6DOF-eq6}) can be modeled by using the maneuvering theory, which requires the following conditions \citep{fossen2011handbook}: first, the zero-frequency wave excitation assumption, which results in the natural periods $T_{n}$ of the ship being in the range of 100-150 seconds; and second, the ship has constant positive linear speed $U = \sqrt{u^2 + v^2 + w^2}$. Under these conditions, the model (\ref{eq:6DOF-eq1})-(\ref{eq:6DOF-eq6}) simplifies as
\begin{align}
m[\dot{u} - vr - x_{g}r^2 - y_{g}\dot{r}] = X \label{eq:CalmWaters1}\\
m[\dot{v} + ur - y_{g}r^2 + x_{g}\dot{r}] = Y \label{eq:CalmWaters2}\\
I_{z}\dot{r} + m[x_{g}(\dot{v} + ur) - y_{g}(\dot{u} - vr)] = N \label{eq:CalmWaters3},
\end{align}
which is commonly used to replace (\ref{eq:6DOF-eq1})-(\ref{eq:6DOF-eq6}) when the above conditions are satisfied such as in sheltered waters or in a harbor. If the condition of calm waters is violated, the seakeeping theory is used, which requires the following conditions: $T_{n} < 100$ and $U$ and $\psi$ are constant, where $\psi$ is the angular position about the $z$-axis \citep{fossen2011handbook}. With these conditions, the system (\ref{eq:6DOF-eq1})-(\ref{eq:6DOF-eq6}) is formulated using the Cummins equation (see \citep{fossen2011handbook} for details). In this example of ship modeling, the model update of a digital twin of the ship occurs whenever there is a switch from the model (\ref{eq:CalmWaters1})-(\ref{eq:CalmWaters3}) to the model based on the seakeeping theory; the model change is implemented in the digital twin to follow the conditions in which the real ship is operating. Fig. \ref{fig:ModelUpdateExample} summarizes this example showing what in this review is considered to be a model update (i.e., Component 3 in Definition 1): changes in the values of the model parameters or variables do not modify the model written in a symbolic form, hence they are classified as data flows between the physical system and the model, i.e., as P $\rightarrow$ M in Fig. \ref{fig:SchemeOfDT}.     

\subsection{Model Update in Statistics-Based Modeling}
The previous subsection covered the case of a digital twin whose model is based on the laws of physics. Another approach is to define the model by defining statistical properties of sensor data coming from the physical system of interest such as in \citep{coraddu2019data} and in \citep{fang2022fatigue}, where a neural network and a Gaussian process are used, respectively. In this review we assume that with statistics-based approaches a model update occurs whenever there is a switch between models based on statistical properties corresponding to different real conditions in the physical system. For example, similarly to the case of the digital twin of a ship discussed in the previous subsection, the model update with statistics-based modeling occurs if there is a switch between a model based on the statistical properties of data recorded with calm waters (i.e., condition A in Fig. \ref{fig:ModelUpdateExample}) and a model based on statistical properties of data recorded with wave excitation (i.e., condition B in Fig. \ref{fig:ModelUpdateExample}). Hence, there is a duality between specific physical conditions and the corresponding statistical properties: new conditions require to update the model through a re-evaluation of its statistical properties. If multiple re-evaluations yielding multiple models do not occur (neither off-line nor on-line), we assume that a model update, i.e., Component 3, is not implemented. Note that, since with machine learning methods the statistical properties are evaluated automatically via training, a model update occurs if there are either multiple off-line trainings for different physical conditions or an on-line training which adapts the model to different conditions.

\section{Results and Analysis}\label{sec:results}
\subsection{Results}
This subsection reports the results of the literature review performed as explained in the previous section. The main points of each work are described below, while Table \ref{tab:OnMarineApplication} summarizes the outcome of the review.

Vasanthan et al. \citep{vasanthan2021combining} developed a path-planning strategy for a vessel comparing different supervised learning methods such as support vector machines \citep{bishop2006pattern} and random forests \citep{breiman2001random}. Their digital twin is a physics-based model used to generate training data for the path-planner; neither data flows nor a model update were implemented. Data flows and model update have not been implemented also in \citep{kutzke2021subsystem} and \citep{lambertini2022underwater}: in the former, a graph of the digital twin of an unmanned underwater vehicle is defined where the nodes are sub-models of the whole model; it is unclear whether the authors used a physics-based or statistics-based modeling approach; in the latter, an underwater drone is designed and built while the digital twin is left as an important future work. To estimate the speed loss of a ship due to marine fouling, Coraddu et al. \citep{coraddu2019data} trained an extreme learning machine \citep{huang2006extreme} on real-world data. The goal of their DT is to advice on the cleaning scheduling of the ship, hence the twin predictions are read by a human rather than directly by the physical system, which means that their data flow is P $\rightarrow$ M $\rightarrow$ H according to our nomenclature. The same data flow was implemented in \citep{taskar2021comparison}, where a physics-based model was used to predict the effect of the waves on the ship performance. Major et al. \citep{major2021real} and VanDerHorn et al. \citep{vanderhorn2022towards} are among the few works that implemented a model update, i.e., Component 3: in the former, the conditions of vessel remote monitoring, vessel maneuvering under harsher weather and for moving the ship crane are considered by switching between different models, whereas in the latter the update of the model is implemented following an assessment of the vessel fatigue damage accumulation. In \citep{uehara2021digital} the derivatives and resistance coefficient were estimated using an extended Kalman filter \citep{ribeiro2004kalman} while the model of the ship was based on the maneuvering theory introduced in the previous section; as the goal of the prediction model is to inform maintenance scheduling, the model output is read by a human rather than fed back directly into the physical system. The zero-frequency waves assumption was adopted also in \citep{rolandsen2018digital} for the digital twin of a vessel moving in ice along with the design of the guidance, navigation and control systems. The authors also considered three different operative conditions of the thrusters based on the vessel speed: below 1.5 m/s, below 3 m/s and above 3 m/s; hence, a model update is implemented in \citep{rolandsen2018digital}. Hence, so far, none of the papers have implemented a bi-directional data flow as required in Component 2.

A bi-directional data flow has been implemented by Fonseca et al. \citep{fonseca2022standards}, where a small-scale hull model was used for the experiments, specifically with a 1:70 scale. A digital twin web-app for monitoring and control of the small-scale hull receives information about the dynamic positioning system and wave characteristics (P $\rightarrow$ M), while the dynamic positioning system on the hull receives the setpoint from the web-app (M $\rightarrow$ P). It is unclear whether the authors used a physics-, a statistics-based modeling approach or a combination of them. In contrast, the seakeeping theory is used in \citep{han2021vessel} since, as pointed out by the authors, a machine learning approach would require a large amount of data. The model update, i.e., Component 3, was not implemented because the paper aim was to update the model parameters using simulated onboard sensor data. The work of \O vereng et al. \citep{overeng2021dynamic} proposes deep reinforcement learning for dynamic positioning, whose neural network was trained using data generated by a physics-based model of the ship. Subsequently, the trained network was implemented on the physical system, which was a 1:20 scale model of a ship. Since the controller was implemented in the real system without the physics-based model, there was no any data flows between P and M in \citep{overeng2021dynamic}. 

All the papers reviewed so far have considered a vehicle or part of it as the physical system. In contrast, \citep{hofmann2019implementation} and \citep{damiani2019digital} targeted a port while \citep{fang2022fatigue} targeted an off-shore platform. These three works have also in common that they have implemented a P $\rightarrow$ M $\rightarrow$ H data flow, hence the human operator was the destination of the model predictions. While the first two papers do not specify the modeling approach used to define Component 1, the third one used Gaussian process regression with real data. Another work that does not mirror a vehicle is \citep{tan2021digital}, where the authors developed a twin of a marine engine room simulator for remote maintenance assistance.        

The thesis \citep{danielsen2018digital} reported several developments towards a digital twin of an electric ship prototype, therefore the author's intent is not to fully implement the twin. In particular, a neural model was trained in a supervised fashion on real data coming from motor states such as speed and current with both faulty and normal conditions in order to detect motor faults. A neural network has been used also by Anyfantis et al. \citep{anyfantis2021abstract} to monitor the condition of ship hulls. The neural model was trained on data generated by a finite-element model. A characteristic of Anyfantis's work is that both physics- and statistics-based modeling approaches were used. Moreover, since monitoring without a feedback loop was the paper goal, an M $\rightarrow$ P data flow was not implemented, whereas there was a P $\rightarrow$ M data flow to optimize the finite-element model. 

A digital twin of a communication network for maritime applications was discussed by Yang at al. \citep{yang2020ai}; since their article is a perspective paper insead of a standard research work, the system is discussed without an implementation. In particular, they introduce the application of artificial intelligence methods and consider a ship as a case study. Cheng et al. \citep{cheng2020digital} are among the few authors that have implemented a bi-directional data flow P $\leftrightarrow$ M. They use a neural network to perform on-line tool life prediction (P $\rightarrow$ M data flow) and a genetic algorithm to optimize the process parameters (M $\rightarrow$ P data flow). Perabo et al. \citep{perabo2020digital} proposed a digital twin based on co-simulation, that is, the simulation of multiple sub-simulators which model different parts of the physical system, e.g. the diesel engine, the electric power plant, the propeller. The sub-simulators are solved independently with their own step sizes and solvers and they are mutually connected to exchange data. According to our review, Wang at al. \citep{wang2019sensor} are the only authors that implemented all the three components of a digital twin as defined in Definition 1. They model the kinematics of a submarine using physics principles along with a stochastic term for the uncertainties. They update on-line the probability density function used to define the submarine states which corresponds to the implementation of Component 3, i.e., a model update. Moreover, as stated by the authors in ``[...] after finding out the best control actions for the anti-submarine ship's second leg heading, we let the anti-submarine ship make its own decision by using the proposed on-line sensor control method. [...]", a feedback loop sends data back to the physical system thus realizing a bi-directional data flow, i.e., Component 2. 

Digital twins of corrosion and crack propagation phenomena are proposed in \citep{wang2021development} and \citep{zhang2019predicting}, respectively: the former predicts the ship corrosion for different sailing routes, while the latter aims at simulating in a laboratory setting several complex behaviors of marine structures such as degradation over time, failures, and redundant load paths. In contrast, parts of marine vehicles are modeled in \citep{bjorum2019development}, \citep{manngaard2020using}, and \citep{johansen2019digital} using a physics-based approach: the first work focused on propulsion systems, the second focused on a heat exchanger for engine cooling and the last focused on drivetrain systems.                            
\begin{table*}[t]
\centering
\caption{Summary of the review outcome. Each entry specifies whether a specific component is implemented or not. The Components are defined in Definition 1 and Fig. \ref{fig:SchemeOfDT}. The modeling approach used for Component 1 and the direction of data flow for Component 2 are between parentheses. In line with Definition 1, the entry for Component 2 is ``Yes'' only if an article explicitly implements a bi-directional data flow between the physical system and the model, i.e., P $\leftrightarrow$ M. The fifth column specifies the type of physical system. The last column is a score that measures how close each article is to a \emph{full implementation} of a digital twin and it is calculated by assigning 1 point to a ``Yes'' entry, 0.5 points to a ``No (P $\rightarrow$ M $\rightarrow$ H)'' entry, and 0 points otherwise. Hence, the maximum score is 3.\\}
\bgroup
\def\arraystretch{1}
\begin{threeparttable}
\begin{tabular}{c@{\hskip3pt}c c c c c} 
 \hline
 Work & Comp. 1 (approach) & Comp. 2 (direction) & Comp. 3 & Ph. system & Score\\  
 \hline\hline
\citep{vasanthan2021combining} & Yes (physics) & No (none) & No & Vessel & 1\\ 
 \hline
\citep{coraddu2019data}  & Yes (statistics) & No (P $\rightarrow$ M $\rightarrow$ H) & No & Ship & 1.5\\ 
 \hline
 \citep{taskar2021comparison} & Yes (physics) & No (P $\rightarrow$ M $\rightarrow$ H) &  No & Ship & 1.5\\ 
 \hline
 \citep{kutzke2021subsystem}  & Yes (n. d.)\tnote{1} & No (none) & No & UUV\tnote{2} & 1\\ 
 \hline
  \citep{rolandsen2018digital} & Yes (physics) & No (P $\rightarrow$ M $\rightarrow$ H) & Yes & Vessel & 2.5\\ 
 \hline
\citep{vanderhorn2022towards} & Yes (both) & No (P $\rightarrow$ M $\rightarrow$ H) &  Yes & Vessel & 2.5\\ 
 \hline
\citep{major2021real} & Yes (physics) & No (P $\rightarrow$ M $\rightarrow$ H) & Yes & Vessel & 2.5\\
  \hline
\citep{uehara2021digital} & Yes (physics) & No (P $\rightarrow$ M $\rightarrow$ H) & No & Ship & 1.5\\
  \hline
\citep{lambertini2022underwater} & No & No (none) & No & Drone & 0\\
  \hline
  \citep{fonseca2022standards} & Yes (n. d.) & Yes & No & Ship & 2\\
  \hline
    \citep{han2021vessel} & Yes (physics) & No (P $\rightarrow$ M $\rightarrow$ H) & No & Vessel & 1.5\\
  \hline
    \citep{overeng2021dynamic} & Yes (physics) & No (none) & No & Vessel & 1\\
  \hline
    \citep{hofmann2019implementation} & Yes (n. d.) & No (P $\rightarrow$ M $\rightarrow$ H) & No & Port & 1.5\\
  \hline
    \citep{damiani2019digital} & Yes (n. d.) & No (P $\rightarrow$ M $\rightarrow$ H) & No & Port & 1.5\\
  \hline
    \citep{fang2022fatigue} & Yes (statistics) & No (P $\rightarrow$ M $\rightarrow$ H) & No & Platform & 1.5\\
  \hline
    \citep{danielsen2018digital} & Yes (statistics) & No (P $\rightarrow$ M $\rightarrow$ H) & No & Ship & 1.5\\
  \hline
    \citep{anyfantis2021abstract} & Yes (both) & No (P $\rightarrow$ M $\rightarrow$ H) & No & Ship & 1.5\\
  \hline
    \citep{tan2021digital} & Yes (n. d.) & No (P $\rightarrow$ M $\rightarrow$ H) & No & MERS\tnote{3} & 1.5\\
  \hline
   \citep{yang2020ai} & No & No (none) & No & \makecell{Communication \\ network} & 0\\
  \hline
  \citep{cheng2020digital} & Yes (statistics) & Yes & n. d. & Engine & 2\\
  \hline
  \citep{perabo2020digital} & Yes (physics) & No (M $\rightarrow$ H) & No & Ship parts & 1\\
  \hline
  \citep{wang2019sensor} & Yes (both) & Yes & Yes & Warfare & 3\\
  \hline
  \citep{bjorum2019development} & Yes (physics) & No (P $\rightarrow$ M $\rightarrow$ H) & No & Vessel & 1.5\\
  \hline
  \citep{zhang2019predicting} & Yes (statistics) & No (P $\rightarrow$ M $\rightarrow$ H) & Yes & Structure & 2.5\\
  \hline
  \citep{wang2021development} & Yes (statistics) & No (P $\rightarrow$ M $\rightarrow$ H) & No & Ship & 1.5\\
  \hline
  \citep{manngaard2020using} & Yes (physics) & No (P $\rightarrow$ M $\rightarrow$ H) & No & Vessel & 1.5\\
  \hline
  \citep{johansen2019digital} & Yes (physics) & No (P $\rightarrow$ M $\rightarrow$ H) & No & Drivetrain & 1.5\\
  \hline
\hline
\end{tabular}
\begin{tablenotes}\footnotesize
\item[1] n. d.: not defined
\item[2] UUV: unmanned underwater vehicle
\item[3] MERS: marine engine room simulator
\end{tablenotes}
\end{threeparttable}
\egroup
\label{tab:OnMarineApplication}
\end{table*}

\subsection{Analysis}
The last column of Table \ref{tab:OnMarineApplication} measures to what extent each reviewed article is close to a \emph{full implementation} of a digital twin, that is, a digital twin with \emph{all the three components} as defined in Definition 1 and Fig. \ref{fig:SchemeOfDT}. As visible, only Wang et al. (\cite{wang2019sensor}) implemented all the three components, hence their score is 3. A score of 2.5 is achieved by Rolandsen and Hoel (\cite{rolandsen2018digital}), VanDerHorn et al. (\cite{vanderhorn2022towards}), Major et al. (\cite{major2021real}), and Zhang and Collette (\cite{zhang2019predicting}). The remaining 22 works (i.e., 81.5\% of the reviewed articles) implemented 2 or less components. In contrast, the physical model (i.e., Component 1) was implemented in 25 articles (i.e., 92.6\% of the reviewed articles). The low number of works with a score higher than 2 (i.e., 5 works) indicates that \emph{full implementations} of digital twins is not yet a common practice. 

Which are the most and the least implemented components? Component 2 is the least implemented (with 3 ``Yes'' entries) closely followed by Component 3 (with 5 ``Yes'' entries). Component 1 is by far the most implemented (with 25 ``Yes'' entries). This result suggests that, to date, Components 2 and 3 are the main challenges to the development of \emph{full} digital twins for marine operations, and hence, they should be the focus of future research.

\section{Conclusion}
Digital twins can provide an effective approach to the optimization of maritime operations. Hence, we reviewed their implementation to understand the current trend in the marine literature and, subsequently, identify the main research gaps. In total, we reviewed 27 articles published since 2018. 

We found that 20 works (i.e., 74.1\%) reached a score less than 2 and only 3.7\% of the works reached the maximum score, meaning that \emph{full implementations} of digital twins are yet uncommon in the literature. In particular, since the physical model is usually implemented (in 92.6\%), the physical model is commonly considered a component of a digital twin. In contrast, the model update and the bi-directional data flow are typically not implemented: missing in 81.5\% of the works the former, in 88.9\% the latter. 

As a consequence of the review findings, future work will focus on implementing \emph{full digital twins}, and, in particular, the model update and the bi-directional data flow.

\section*{Acknowledgements} 
This work was supported by the Natural Environment Research Council, United Kingdom [grant number \\ NE/V008080/1].

\bibliography{mybibfile}

\end{document}